# Detection of Sugar-Lectin Interactions by Multivalent Dendritic Sugar Functionalized Single-Walled Carbon Nanotubes

K. S. Vasu[1], K. Naresh[2], R. S. Bagul[2], N. Jayaraman[2] and A. K. Sood[1,*]

[1]Department of Physics, Indian Institute of Science, Bangalore-560012, India.

[2]Department of Organic Chemistry, Indian Institute of Science, Bangalore-560012, India.

**Abstract:**

We show that single walled carbon nanotubes (SWNT) decorated with sugar functionalized poly (propyl ether imine) (PETIM) dendrimer is a very sensitive platform to quantitatively detect carbohydrate recognizing proteins, namely, lectins. The changes in electrical conductivity of SWNT in field effect transistor device due to carbohydrate – protein interactions form the basis of present study. The mannose sugar attached PETIM dendrimers undergo charge – transfer interactions with the SWNT. The changes in the conductance of the dendritic sugar functionalized SWNT after addition of lectins in varying concentrations were found to follow the Langmuir type isotherm, giving the concanavalin A (Con A) – mannose affinity constant to be $8.5 \times 10^6$ $M^{-1}$. The increase in the device conductance observed after adding 10 nM of Con A is same as after adding 20 μM of a non – specific lectin peanut agglutinin, showing the high specificity of the Con A – mannose interactions. The specificity of sugar-lectin interactions was characterized further by observing significant shifts in Raman modes of the SWNT.

Probing carbohydrate–protein interactions is important in establishing the roles of cell surface carbohydrates which are essential for many cascade events happening on the cell walls [1]. Carbohydrate-binding proteins, namely, lectins bind to carbohydrate ligands with weak binding affinities, ranging in molar to sub-millimolar concentrations, even when preserving high specificities of the interaction. Nature overcomes barriers of weak binding affinities through the so-called multivalency, wherein clustered patches of sugars enhance lectin binding affinities exponentially, so as to make the interactions physiologically relevant [2]. Agglutination is one of the early methods to monitor sugar-receptor interactions [3], leading to a variety of biological and biophysical assays [4]. Whereas prominent hemagglutination assays rely on the basis of kinetically driven aggregations, thermodynamic characterizations involve the primary utilization of calorimetric methods to follow the interactions occurring in homogeneous solutions [5]. Other methods to map the interactions, closely matching that of two-dimensional cell surfaces, are the Langmuir [6], the surface plasmon resonance (SPR) [7] and quartz crystal microbalance (QCM) methods [8, 9] that allow the characterization of kinetic parameters, in addition to deriving equilibrium binding constants. Fluorescence based assays were also used in cases when either a fluorophore or chromophore engages in a reporter role of the recognition events [10]. Low binding affinities, coupled with kinetically-driven aggregations and/or thermodynamically driven chelations of ligand-receptor complexes, result in varied levels of detection limits. X – ray crystallography and nuclear magnetic resonance (NMR) techniques have been used to investigate carbohydrate–protein interactions by determining the structure of proteins and their biological functions [11-14]. In electronic impedance spectroscopy method, the carbohydrate modified boron doped diamond surfaces have been used to investigate carbohydrate–protein interactions [15]. Due to the excellent electronic properties of the SWNTs, the field effect transistor (FET) made of SWNTs and chemically modified SWNTs were used as protein biosensors [16, 17]. FETs made of biotin attached SWNTS and ZnS nanocrystals decorated SWNTs were used to detect label free protein and DNA respectively [18, 19]. SWNTs based chemi-resistive FET sensors were used to detect carbohydrates (glucose, fructose) with Con A, where SWNTs were incubated in dextran that attaches Con A to SWNT with low affinity [20]. Recently liquid gated FETs made of porphyrin based glycoconjugates attached to SWNTs were used to probe the interactions between carbohydrates and their specific proteins [21].

In exploring the possibilities to interface characteristic electrical properties of SWNTs, the significant functionalization of SWNT with multivalent presentation of ligands or receptors will enhance the sensitivity to probe the specific carbohydrate–protein interactions. In this paper we present dendritic macromolecules as platforms for multivalent presentations of carbohydrates and the functionalization of SWNT with these dendritic molecules in order to probe the carbohydrate-protein interactions. For this purpose, multivalent mannose sugar–coated PETIM dendrimer [22, 23] and mannose–specific lectin Con A were chosen to study the interactions by analyzing the changes in the electrical conductivity of the FET device made of SWNTs. Analysis showed that changes in the conductivity of FET device could be fitted to a Langmuir type isotherm to extract equilibrium affinity constant of the carbohydrate–protein interactions. In addition, the interactions occurring over the surface of dendritic sugar functionalized SWNT were also followed by changes in the Raman spectra of the SWNT resulting from charge transfer interactions.

SWNTs having an average diameter of 0.8 nm, predominantly semiconducting and (6, 5) chirality with purity of 90%, were obtained from M/S South West Nano Technologies. Source and drain electrodes with separation of 5 μm were patterned on 300 nm $SiO_2$ /Si using photolithography (MicroTech Laser Writer LW 405), followed by RF sputtering of 10 nm Cr and 40 nm Au. Devices were made by using ac dielectrophoresis (10 V peak to peak at 1 MHz) of 3 μL of SWNTs dispersion (made by dispersing 10 μg of pristine nanotubes in 5 mL of 1, 2-dichloroethane and sonicating for 3 hr) in-between two patterned Cr/Au electrodes. WITEC confocal Raman spectrometer and 514 nm laser radiation from an argon ion laser with power less than 1 mW, were used to perform Raman measurements.

Preparation of mannose derivatized dendrimer (DM): Tetra-O-benzoyl mannosyl bromide (0.38 g, 0.59 mmol) in $CH_2Cl_2$ (30 mL) was added to a solution of dendritic alcohol (G4-OH) [23] (0.048 g, 9.2 µmol) and $Ag_2CO_3$ (0.16 g, 0.59 mmol), stirred for 36 h at room temperature, filtered and the filtrate concentrated *in vacuo* and purified by column chromatography ($Al_2O_3$, EtOAc:MeOH = 9:1). The resulting product in tetrahydrofuran (THF) : MeOH (1:1) (10 mL) was added with NaOMe in MeOH (1 M) (0.01 mL), stirred at room temperature for 12 h, neutralized with Amberlite ion-exchange ($H^+$) resin, filtered, concentrated *in vacuo* and the resulting solution subjected to dialysis (MW cut-off 3.5 KD) to afford DM as a foamy solid. Yield: 0.035 g, (39 %, after two steps); $^1$H NMR ($D_2O$, 400 MHz) δ 1.76 (br s, 180 H), 2.71 (br s, 180 H), 3.48-3.59 (m, 200 H), 3.68 (br s, 110 H), 3.79-3.85 (m, 70 H), 4.77 (s, 30 H); $^{13}$C NMR ($D_2O$, 100 MHz) δ 24.7, 24.8, 50.2, 61.0, 65.5, 66.7, 66.8, 68.5, 70.1, 70.7, 72.9, 99.8. A comparison of $^1$H NMR integrations of 4.77 ppm (H-1 of sugar) and 1.87 ppm (-$CH_2$-C*H$_2$*-$CH_2$-) of dendrimer structure showed about 30 sugar units functionalizing a dendrimer molecule. Phenol-sulfuric acid assay [24] further confirmed the extent of mannose functionalization as deduced by $^1$H NMR spectroscopy. Mannopyranoside-derivatized dendrimer (DM) was soluble in water and a solution of concentration 0.44 mM was used for the studies. Solutions of Con A and peanut agglutinin (PNA) were prepared in distilled water (pH at 6.5 – 6.7) in the concentration range of 1 nM to 20 μM.

Inset of Fig. 1a shows the schematic of a FET device to probe DM–Con A interactions and the molecular structure of DM. Fig. 1a also shows the source-drain current-voltage characteristics ($I_{DS}$-$V_{DS}$) of SWNT alone and DM functionalized SWNT. The resistance of the SWNT device increased ~16 times upon addition of 2 μL of 0.44 mM dendrimer solution. The transfer characteristics, source-drain current ($I_{DS}$) as a function of back gate voltage ($V_{BG}$) of pristine SWNT at a constant source-drain voltage $V_{DS}$ of 0.2 V shown in Fig. 1b, establish that the SWNTs are unintentionally hole doped. On addition of 2 and 4 μL of DM solutions, the source – drain current decreases and the charge neutrality point shifts towards negative gate voltages, indicating a charge transfer type interaction between the SWNT and the dendrimer, similar to the previous studies of hydroxyl-group terminated PETIM dendrimer-SWNT complexation occurring through multitude lone pairs of electrons arising from oxygen and nitrogen atoms constituting the dendrimer [25]. The interaction resulted in a reduction of the effective carrier density of hole doped SWNT due to electron-doping and generation of more scattering centers on the SWNT surface, resulting in the increase of the resistivity of the device. Inset of Fig. 1b shows the hysteresis observed in the FET device made of pristine SWNT and after adding 2 μL of DM, attributed to trapped charges [26].

Lectin binding studies were performed using aqueous lectin solutions by adding each time 2 μL of lectin solutions, in the concentrations ranging from 1 nM to 20 μM. Fig. 2 shows the ratio $\Delta G/G_0$ of the change in conductance of the device (at zero gate voltage) upon addition of varying concentrations of lectin solutions, where $G_0$ is the conductance of the SWNT + DM and $\Delta G$ is G(SWNT + DM + lectin) – $G_0$(SWNT + DM). We note that $\Delta G/G_0$ ~ 3% even for addition of 1 nM Con A, pointing out that sensitivity of detection of Con A is atleast 1 nM. Comparing our results with Ref. [21] where mannose attached porphyrin based glycoconjugates were used, the change $\Delta G/G_0$ for addition of 2 μM of Con A in the present work is ~ 40 times larger than the $\Delta G/G_0$ for addition of 2 μM of Con A in Ref. [21]. The conductance of the device is seen to increase with the concentration of Con A due to a net positive charge on Con A (at pH 6.5 -6.7) which induces hole doping in the nanotubes. This is further corroborated by the Raman spectra of SWNTs (to be discussed later). In case of non-specific interaction of lectin PNA with mannose, the increase in the electrical conductance of SWNT + DM was an order of magnitude less. In other words, the change in $\Delta G/G_0$ for 20 μM of PNA is same as for 10 nM of Con A.

The change in conductance as a function of lectin concentration follows the Langmuir type adsorption isotherm [27],

$$\frac{\Delta G}{G_0} = S[\log C + \log(7.389 K_A)] \qquad (1)$$

where the parameter S is called structure factor, C is the concentration of Con A solution and $K_A$ is the affinity constant. In Fig. 2 the plot of $\Delta G/G_0$ (solid points) for Con A lectin as function of C is fitted to Eq. (1), giving $K_A = 8.5 \times 10^6$ M$^{-1}$. To observe the effect of gate voltage on the sensitivity of detection, experiments were also performed at a backgate voltage $V_{BG}$ = +5 V. Inset of Fig. 2 shows change in conductance of the device upon addition of varied amounts of lectin at $V_{BG}$ = 0 V and +5 V. An enhancement in $\Delta G/G_0$ is seen at positive gate voltages, which corresponds to electron doping, showing the enhancement in the sensitivity.

Interactions between the lectins and DM-functionalized SWNT were also monitored using tangential Raman bands of the SWNT as these are known to be sensitive to electronic perturbations [28]. Fig. 3 shows Raman spectrum of pristine SWNT displaying longitudinal optic mode (LO, $G^+$-mode), transverse optic mode (TO, $G^-$-mode) and defect induced D-mode. Spectra are also shown after adding DM and then 2 µL of 20 µM Con A to the SWNT + DM complex. The solid lines in Fig. 3 show the Lorentzian fits to the data represented by open circles. The frequencies, line widths and area ratios of the Raman modes are given in Table I. It is seen that addition of DM leads to a red-shift of 6.3 cm$^{-1}$ in $G^-$ mode of SWNT and the line width increased from 50.3 cm$^{-1}$ to 62.5 cm$^{-1}$. Following quantitative study of the effect of doping on $G^-$ and $G^+$ modes of semiconducting nanotubes [28], the observed shift in $G^-$ mode is attributed to the decrease in hole concentration of SWNT due to the electron transfer after addition of DM. On further adding Con A to SWNT + DM complex, $G^-$ mode is blue-shifted by 12.4 cm$^{-1}$ which is due to hole doping in the nanotube [28], thus confirming our earlier reasoning for the increase in the conductivity of SWNT + DM complex on addition of Con A. Also the increase in the ratio $A_D / A_{G+}$ suggests the increase in the defect concentration in SWNTs after adding DM.

We have studied binding abilities of sugar coated PETIM dendrimers with the SWNT to explore read-out capabilities of SWNT as applicable to sugar-lectin interactions. Fourth generation PETIM dendrimer was functionalized covalently with mannopyranoside residues at their peripheries to get multivalent sugar coated dendrimer which interacts with the SWNT through charge-transfer interactions. Upon ensuring the interaction of multivalent sugar coated dendrimer with the SWNT, the conductivity of the SWNTs was used to assess carbohydrate-protein interactions, using lectin Con A with known binding sites for mannose and non-specific lectin PNA. Interactions with the lectin Con A result in increase in the conductance of the SWNT + DM complex and differential change in the conductance can be fitted to Langmuir type isotherm, yielding a quantitative measure of affinity constant to be $8.5 \times 10^6$ $M^{-1}$. The selectivity of change in conductance was larger for lectin Con A as compared with lectin PNA. Such a large change in conductance as a result of interaction with lectins, differentiated by specificities of sugar-lectin interactions, is by far un-known. The observed $\Delta G/G_0$ in our experiments is ~ 40 times higher than that of Ref. [21], shows that the multivalent mannose decorated dendrimer is more sensitive than mannose attached porphyrin based glycoconjugates. Raman spectral studies further confirmed the read-out capabilities of dendrimer functionalized SWNT upon interaction with lectins, through large shifts in the tangential Raman modes of the SWNT. The present study establishes that dendrimer functionalized SWNT is a valuable tool to follow carbohydrate-protein interactions, with sensitivities achievable using lectin solutions as low as 1 nM. The method opens up newer possibilities of dendrimer-SWNT devices interfacing biologically relevant recognitions.

We thank Department of Science and Technology, India, for financial support under the Nanomission grant.

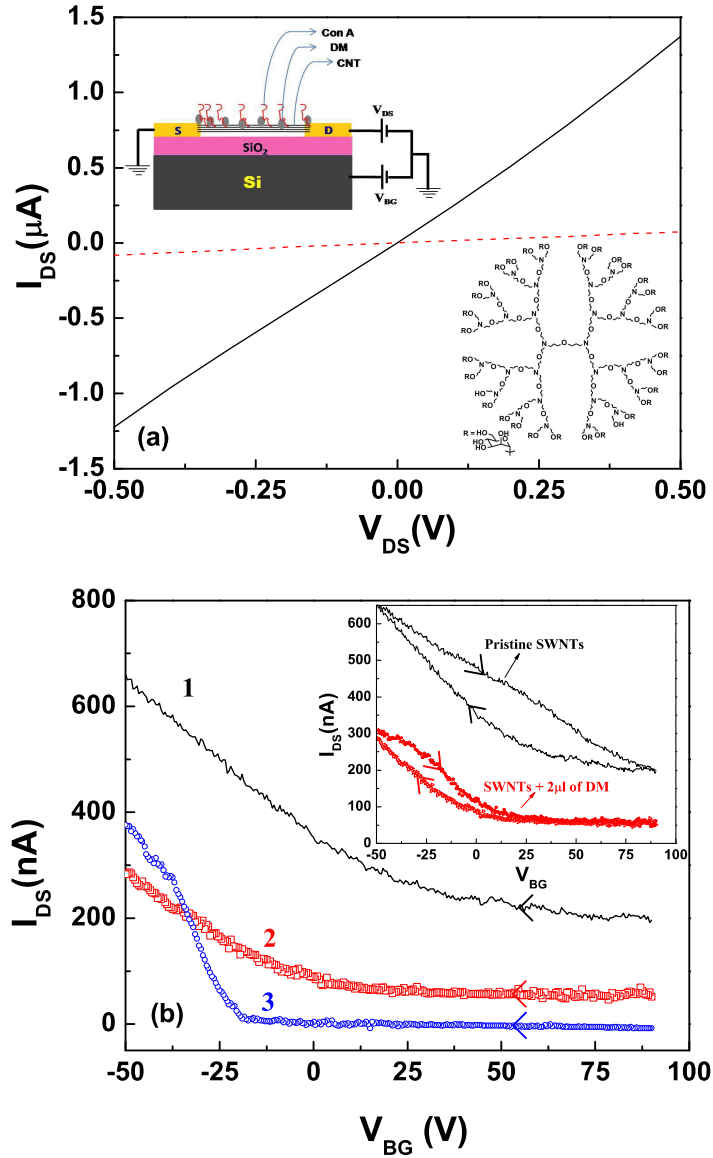

Fig.1. (a) Source – drain current $I_{DS}$ – $V_{DS}$ characteristics of SWNTs (solid line) and SWNTs + 2μL of DM (dash line). Inset: Schematic of the FET device and the molecular structure of mannose decorated fourth generation PETIM dendrimer. (b) Transfer characteristics, $I_{DS}$ as function of back gate voltage ($V_{BG}$) of, (1) Pristine SWNT (2) 2 μL and (3) 4 μL of DM when the gate voltage is ramped from +90 V to -50 V. The inset shows the hysteresis that observed in the device made of pristine SWNT and SWNT + 2 μL of DM.

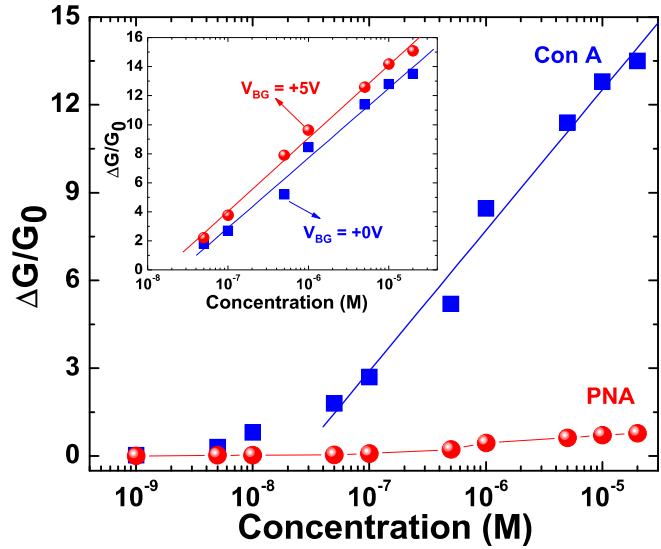

Fig.2. $\Delta G/G_0$ of the device (a) after adding 2 μL of lectin solution of different concentrations on the SWNT + DM complex. Inset: $\Delta G/G_0$ vs concentration of con A, fitted using Eq. (1), at $V_{BG} = 0$ and $V_{BG} = +5$ V.

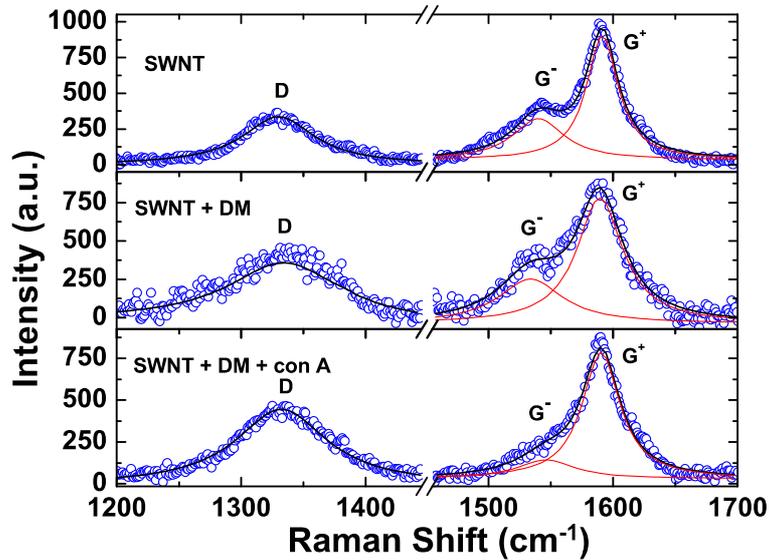

Fig.3. Defect induced mode (D - mode), transverse optic mode (TO, $G^-$ - mode) and longitudinal optic mode (LO, $G^+$ - mode) of pristine SWNTs, SWNTs + DM and SWNTs + DM + 2 μL of 20 μM con A recorded using an excitation energy of 2.41eV. (Blue color open circles show the raw Raman spectra and the lines are the fitted one).

**Table I:** The frequencies, line widths of the Raman modes ($G^+$, $G^-$, and D) of SWNT, SWNT + DM, SWNT + DM + Con A.

| | $G^+$ | | $G^-$ | | D | | $A_D/A_{G+}$ |
|---|---|---|---|---|---|---|---|
| | $\omega$ (cm$^{-1}$) | $\Gamma$ (cm$^{-1}$) | $\omega$ (cm$^{-1}$) | $\Gamma$ (cm$^{-1}$) | $\omega$ (cm$^{-1}$) | $\Gamma$ (cm$^{-1}$) | |
| SWNT | 1591±0.7 | 30.9±2.4 | 1540±0.4 | 50.3±4.1 | 1330.3±0.6 | 70.2±5.4 | 0.9 |
| SWNT+DM | 1589.4±0.5 | 44.8±1.7 | 1533.7±1.8 | 62.5±6.1 | 1334±0.8 | 102.3±3.2 | 1.1 |
| SWNT+DM+Con A | 1590.4±0.3 | 36.8±0.9 | 1546.1±2.9 | 55.8±7.4 | 1332±1.6 | 89.3±2.2 | 1.4 |

**References:**

(1) N. Sharon, H. Lis, Science **246**, 227 (1989).

(2) M. Mammen. S.-K. Choi, and G.M. Whitesides, Angew. Chem. Int. Ed. **37**, 2754 (1998).

(3) L. L. So, and I. J. Goldstein, J. Biol Chem. **242**, 1617 (1967).

(4) Y. C. Lee and R. T. Lee, Neoglycoconjugates: preparation and applications Academic Press, San Diego, 23 (1994).

(5) E. J. Toone, Curr. Opin. Struct. Biol. **4**, 719 (1994).

(6) B. N. Murthy, S. Sampath, and N. Jayaraman, Langmuir **21**, 9591 (2005).

(7) E. A. Smith, W.D. Thomas, L. L. Kiessling, and R. M. Corn, J. Am. Chem. Soc. **125**, 6140 (2003).

(8) Y. Ebara, and Y. A. Okahata J. Am. Chem. Soc. **116**, 11209 (1994).

(9) Y. Zhang, S. Luo, Y. Tang, L. Yu, K.-Y. Hou, J.-P. Cheng, X. Zeng, and P. G. Wang, Anal. Chem. **78**, 2001 (2006).

(10) D. H. Charych, J. O. Nagy, W. Spevak, and M. D. Bednarski, Science **261**, 585 (1993).

(11) F. A. Quiocho, Pure & Appl. Chem. **61**, 1293 (1989).

(12) W. I. Weis and K. Drickamer, Annu. Rev. Biochem. **65**, 441 (1996).

(13) T. K. Dam and C. F. Brewer, Chem. Rev. **102**, 387 (2002).

(14) N. Jayaraman, Chem. Soc. Rev. **38**, 3463 (2009).

(15) S. Szunerits, J. N. Jonsson, R. Boukherroub, P. Woisel, J. S. Baumann, and A. Siriwardena, Anal. Chem. **82**, 8203 (2010).

(16) J. N. Tey, I. P. M. Wijaya, Z. Wang, W. H. Goh, A. Palaniappan, S. G. Mhaisalkar, I. Rodriguez, S. Dunham, and J. A. Rogers, Appl. Phys. Lett. **94**, 013107 (2009).

(17) I. P. M. Wijaya, S. Gandhi, T. J. Nie, N. Wangoo, I. Rodriguez, G. Shekhawat, C. R. Suri, and S. G. Mhaisalkar, Appl. Phys. Lett. **95**, 073704 (2009).

(18) P. Hu, A. Fasoli, J. Park, Y. Choi, P. Estrela, S. L. Maeng, W. I. Milne, and A. C. Ferrari, J. Appl. Phys. **104**, 074310 (2008).

(19) Rajesh, Basanta K. Das, Sira Srinives, and Ashok Mulchandani, Appl. Phys. Lett. **98**, 013701 (2011).

(20) L. N. Cella, W. Chen, N. V. Myung, and A. Mulchandani, J. Am. Chem. Soc. **132**, 5024 (2010).

(21) H. Vedala, Y. Chen, S. Cecioni, A. Imberty, S. Vidal, and A. Star, Nano Lett. **11**, 170 (2011).

(22) T. R. Krishna, and N. Jayaraman, J. Org. Chem. **68**, 9694 (2003).

(23) G. Jayamurugan, and N. Jayaraman, Tetrahedron **62**, 9582 (2006).

(24) M. Monsigny, C. Petit and A.-C. Roche, Anal. Biochem. **175**, 525 (1988).

(25) G. Jayamurugan, K. S. Vasu, Y. B. R. D. Rajesh, S. Kumar, V. Vasumathi, P. K. Maiti, A. K. Sood, and N. Jayaraman, J. Chem. Phys. **134**, 104507 (2011).

(26) W. Kim, A. Javey, O. Vermesh, O. Wang, Y.M. Li, and H. J. Dai, Nano Lett. **3**, 193 (2003).

(27) M. Lee, J. Lee, T.H. Kim, H. Lee, B. Y. Lee, J. Park, Y.M. Jhon, M.-J. Seong, and S. Hong, Nanotechnology, **21**, 055504 (2010).

(28) A. Das and A.K. Sood, Phys. Rev. B **79**, 235429 (2009).